\documentstyle[preprint,aps,a4,12pt]{revtex}
\tightenlines
\begin{document}
\draft
\hfill\vbox{\baselineskip14pt
            \hbox{\bf KEK-TH-525}
            \hbox{KEK Preprint 97-94}
            \hbox{}}
\baselineskip20pt
\vskip 0.2cm 
\begin{center}
{\Large\bf Model for Trapped Ion interacting with Standing Wave}
\end{center} 
\vskip 0.2cm 
\begin{center}
\large S.~Alam\footnote{email:~sher@theory.kek.jp}and 
C.~Bentley\footnote{Prairie View Texas A \& M University, Texas, USA.}
\end{center}
\begin{center}
{\it Theory Group, KEK, Tsukuba, Ibaraki 305, Japan }
\end{center}
\vskip 0.2cm 
\begin{center} 
\large Abstract
\end{center}
\begin{center}
\begin{minipage}{14cm}
\baselineskip=18pt
\noindent
We treat the interaction of a trapped ion with a standing-wave
light field relaxing the previous condition of the trapping ion being
on the node of the standing wave. In addition we work to
second order in the Lamb-Dicke parameter $\eta$, the extension
to higher orders in $\eta$ is also indicated. Only a
simple transformation to the ``dressed basis'' [subject
to the constraint of Pauli algebra] is required
in order to show that the ion interacting with a standing 
light-wave may be described by the Jaynes-Cummings Hamiltonian.
\end{minipage}
\end{center}
\vfill
\baselineskip=20pt
\normalsize
\newpage
\setcounter{page}{2}
	Field theories in particle physics of spin one-half
fermions [matter fields] interacting with spin-one vector bosons
[force fields] subject to the principle of local gauge invariance 
have been very successful in fundamental processes. The familiar
examples are quantum electrodynamics [QED], quantum chromodynamics
[QCD] and quantum flavorodynamics [QFD]. QED the theory of
electrically charged spin one-half fermions interacting
with massless spin one photons is remarkably precise in
predicting physical quantities up to 15 decimal places.
The quantum corrections up to several loop orders in QED 
 precisely and beautifully fit the experimental data.

	In the context of quantum optics the Jaynes-Cummings
model is a favorite among theorists and experimentalists. 
As is well-known when we consider a quantum two-level 
system interacting with a single mode of quantized field one 
is led to the familiar Jaynes-Cummings Hamiltonian [JCH] 
\cite{jay63} provided one is interested only in the difference
of the population of the two levels. The JCH has been extensively 
used as a model Hamiltonian in fields such as 
quantum optics, nuclear magnetic resonance, and quantum 
electronics. It is {\em well-known} that a two-level
system interacting with a single mode of radiation
is mathematically equivalent to spin one-half system
interacting with the a single mode of radiation.
It appears that of authors Yang et al., \cite{yan97} 
seem to overlook
the fact that this similarity between JC model and
one describing a spin-$\frac{1}{2}$ particle
in a magnetic field is well-known, to quote them \cite{yan97}
``We now establish the similarity between JC model
and one describing a spin-$\frac{1}{2}$ particle
in a magnetic field.'' A compact method/notation to 
solve various nonlinear extensions of the 
JCH will be reported elsewhere \cite{ala97-1}.

An interesting study
of the JCH is the periodic spontaneous collapse and revival 
due to quantum granularity of the field \cite{mey90}.
In the rotating wave approximation [RWA], the JCH
becomes solvable and it has been broadly used in the last years
\cite{eber80,rr89,hong92,ho92,mill92,pran92}. There are
many other interesting works on the JCH.

	A question that naturally arises is, why is 
the JCH so successful and rich in predictions?
An answer could be that it is the simplest Hamiltonian
one can write for spin-half system interacting
with a single mode of quantized field and thus
covers a whole spectrum of natural phenomena,
since as already mentioned it seems that we
can formulate a large class of successful
quantum theories based on the interaction
of spin one-half fields interacting with
spin one fields. Exploiting the general
structure of spin one-half system interacting
with radiation field by using well-known results  
a compact method/notation has been developed to 
solve various nonlinear extensions of the 
JCH \cite{ala97-1}. For example in the present note
by demanding that the transformed dressed matrices
obey the Pauli algebra we are able to demonstrate
the correspondence of trapped ion dynamics
interacting with a standing-wave light field
to the JCH.

	One can also attempt to solve Hamiltonians
using the mathematics of Supersymmetry [SUSY].
Although SUSY still awaits experimental verification
in the context of particle physics nevertheless one is 
free to use SUSY as a mathematical tool to solve various 
physical problems. In particular SUSY quantum mechanics
has proved quite useful in the solution of various
problems, see \cite{lah90} for a review. One 
example of SUSY quantum mechanics at work is 
the exact supersymmetry in nonrelativistic hydrogen atom 
considered by Tangerman and Tjon \cite{lah90}, there are 
many others. Solutions of the JCH and its extensions
using SUSY will be reported elsewhere \cite{ala97-2}.

	The Hamiltonian of the trapped ion interacting
with a standing wave where the trapping ion is on the 
{\em node} 
of the standing wave to first order in the Lamb-Dicke 
parameter $\eta$ was given by Cirac et al.,.\cite{cir93}. 
It was shown in \cite{cir93} that the dynamics of the 
laser cooled two-level trapped ion is described by
the JCH. The analogy established in \cite{cir93}
between trapped ion dynamics and cavity QED is
an important one and has resulted in many proposals/ideas
for example quantum non-demolition measurement of final 
temperature \cite{cir94,hel95}, experimental study of 
collapse and revivals \cite{mee96} and theoretical
studies of the same \cite{cir92,cir93,mar94}.
One intuitively expects that the dynamics of the ion
away from the node will also be described by JCH.
The main purpose of this note is to extend the result
of \cite{cir93} by allowing the trapped ion to be anywhere
[not just the node] on the standing-wave light field.
Recently Wu and Yang \cite{wu97} [see Note added]
have also attempted to extend the work of
\cite{cir93}. In their analysis \cite{wu97}
they introduce the interaction picture which
is not necessary as is clear from our work.
More precisely the use of interaction picture in our approach
is not needed in the context of establishing a correspondence
between trapped two-level ion interacting with a standing-wave
light field and JCH. 
One simply needs to use the transformation
to the dressed basis to establish the correspondence
between the trapped ion interacting with
the standing-wave light field and JCH. The
particular form of the transformation is determined
by imposing the condition that algebra of Pauli
matrices be maintained.
Moreover we go beyond the first order in $\eta$
and point out that by keeping second
and higher order terms in $\eta$ we are lead to the
multi-photon or m-photon 
JCH \cite{buck81,suk84,bar91,ala96-1}.
The master equation of a single two-level ion trapped
in a harmonic potential and located at the node
of standing-wave light field is \cite{cir93},
\begin{eqnarray}
\frac{d\rho}{dt}&=&-i\left[\nu a^{\dagger}a+\frac{\Delta}{2}
\sigma_{z}-\frac{\Omega_0}{2}(\sigma_{+}+\sigma_{-})
\sin[\eta(a+a^{\dagger})], \rho\right]\nonumber\\
&&+\frac{\Gamma}{2}(2\sigma_{-}\tilde{\rho}\sigma_{+}
-\sigma_{+}\sigma_{-}\rho-\rho\sigma_{+}\sigma_{-}),
\label{F1}
\end{eqnarray}
where $\Delta$ is the detuning of the two-level transition
from the laser frequency, $\nu$ is the trap frequency,
$\Omega_{0}$ is the laser Rabi frequency, and 
$\eta=\pi a_{0}/\lambda$ is the Lamb-Dicke parameter
with $a_0$ the amplitude of the ground state of the trap 
and $\lambda$ the optical wavelength.
$a^{\dagger}$ and $a$ are the creation and destruction
operators for phonons respectively. The 
$\sigma_{i}$'s $i=x,y,z$ are usual
Pauli matrices \cite{mey90} with $\sigma_{\pm}=\frac{1}{2}
(\sigma_{x}\pm i \sigma_{y})$, the raising and lowering
operators for the two-level system. We note the 
familiar relations governing $\sigma$'s i.e.
$\{\sigma_{i},\sigma_{j}\}=\delta_{ij}$, 
$[\sigma_{i},\sigma_{j}]=2i\epsilon_{ijk}\sigma_{k}$,
$i,\;j=x,\;y,\;z$.
$\tilde{\rho}$
does not concern us here and is defined in \cite{cir93}.
As mentioned in \cite{cir93} it accounts for the 
momentum transfer associated
with spontaneous emission of a photon. 

The Hamiltonian for the single two-level ion trapped in
a harmonic potential and located at the node of a standing-wave
light field is easily read from Eq.~\ref{F1}, one has
\begin {equation}
{\cal H}=\hbar \nu a^{\dagger}a+\hbar \frac{\Delta}{2}
\sigma_{z}-\hbar\frac{\Omega_0}{2}(\sigma_{+}+\sigma_{-})
\sin[\eta(a+a^{\dagger})].
\label{F2} 
\end{equation}
We note that the population occupation operators
are defined in the usual manner 
$\sigma_{22}-\sigma_{11}=\sigma_{z}$ and 
$\sigma_{22}+\sigma_{11}=1$ or alternatively
$\sigma_{22}=\frac{1}{2}(1+\sigma_{z})$, and
$\sigma_{11}=\frac{1}{2}(1-\sigma_{z})$.

It is straightforward to extend Eq.~\ref{F2} to the case
when the ion is not necessarily on the node by letting 
$-\sin[\eta(a+a^{\dagger})]\rightarrow 
-\sin[{\pi/2}]\sin[\eta(a+a^{\dagger})]\rightarrow 
\cos[\eta(a+a^{\dagger})+\pi/2]\rightarrow 
\cos[\eta(a+a^{\dagger})+\chi]$. We thus obtain
the Hamiltonian for the trapped ion in a harmonic 
potential and located anywhere on a standing-wave
light field,
\begin {equation}
{\cal H}=\hbar \nu a^{\dagger}a+\hbar \frac{\Delta}{2}
\sigma_{z}+\hbar\frac{\Omega_0}{2}(\sigma_{+}+\sigma_{-})
\cos[\eta(a+a^{\dagger})+\chi],
\label{F3} 
\end{equation}
We can write $\cos[\eta(a+a^{\dagger})+\chi]=\cos[\eta(a+a^{\dagger})]
\cos{\chi}-\sin[\eta(a+a^{\dagger})]\sin{\chi}$ and expand
Eq.~\ref{F4} in powers of the Dicke parameter $\eta$ 
to obtain,
\begin {eqnarray}
{\cal H}&=&\hbar \nu a^{\dagger}a+\hbar \frac{\Delta}{2}\sigma_{z}
\nonumber\\
&&+\hbar\frac{\Omega_0}{2}(\cos{\chi})(\sigma_{+}+\sigma_{-})
-\eta\hbar\frac{\Omega_0}{2}(\sin{\chi})
(\sigma_{+}+\sigma_{-})(a+a^{\dagger})\nonumber\\
&&-\frac{\eta^2}{2}\hbar\frac{\Omega_0}{2}(\cos{\chi})
(\sigma_{+}+\sigma_{-})(a+a^{\dagger})^2+{\rm O}(\eta^3).
\label{F4}
\end{eqnarray}
For the moment ignoring the term of order $\eta^2$
and using the definition/relation of 
$\sigma_{+}+\sigma_{-}=\sigma_{x}$ we may rewrite
Eq.~\ref{F4} as
\begin {eqnarray}
{\cal H}&=&\hbar \nu a^{\dagger}a+\hbar \frac{\Delta}{2}\sigma_{z}
+\hbar\frac{\Omega_0}{2}(\cos{\chi})\sigma_{x}\nonumber\\
&&-\eta\hbar\frac{\Omega_0}{2}(\sin{\chi})
(\sigma_{+}a+\sigma_{-}a^{\dagger}),
\label{F5}
\end{eqnarray}
where in the last line we have gone to the usual RWA.
We have gone to the RWA since we want to compare
our Hamiltonian to the JCH which is usually written
using RWA.
In order to cast Eq.~\ref{F5} into the JC form we can
combine the second and third terms in Eq.~\ref{F5} into a
single term by defining the transformation
\begin {eqnarray}
\Sigma_{z}&=&\frac{\Delta}{\Omega}\sigma_{z}
+\frac{\Omega_0\cos\chi}{\Omega}\sigma_x,\nonumber\\
\Omega &=&\sqrt{\Delta^2+\Omega_0^2\cos^2\chi}.
\label{F6}
\end{eqnarray}
We note that $\Sigma_z^2=1$. Using \ref{F6} in
Eq.~\ref{F5} we can write the latter as
\begin {eqnarray}
{\cal H}&=&\hbar \nu a^{\dagger}a
+\hbar \frac{\Omega}{2}\Sigma_{z}\nonumber\\
&&-\eta\hbar\frac{\Omega_0}{2}\sin\chi
(\sigma_{+}a+\sigma_{-}a^{\dagger}),
\label{F7}
\end{eqnarray}
In order to have a consistent Pauli algebra
between the $\Sigma$ matrices we must supplement
$\Sigma_z$ in \ref{F6} by $\Sigma_x$ and
$\Sigma_y$ where 
\begin {eqnarray}
\Sigma_{x}&=&\frac{\Delta}{\Omega}\sigma_{x}
-\frac{\Omega_0\cos\chi}{\Omega}\sigma_{z},\nonumber\\
\Sigma_{y}&=& \sigma_{y},\nonumber\\
\Sigma_{\pm}&=& \frac{1}{2}(\Sigma_{x}\pm i \Sigma_{y}),
\label{F8}
\end{eqnarray}
which are/[can be] found/determined after some simple algebra.
For example a simple way to determine $\Sigma_{x}$ is to assume
that it is some arbitrary linear combination of 
$\sigma_{x}$ and $\sigma_{z}$ and then determine the
coefficients appearing in the linear combination
by using the fact that $\{\Sigma_z,\Sigma_x\}= 0$.
It is easily checked that $\Sigma_x^2=1$, $\Sigma_y^2=1$,
$\{\Sigma_x,\Sigma_y\}= 0 $, $\{\Sigma_y,\Sigma_z\}= 0 $
$\{\Sigma_z,\Sigma_x\}= 0 $, $[\Sigma_x,\Sigma_y]= 2i\Sigma_z$, 
 $[\Sigma_y,\Sigma_z]= 2i\Sigma_x$, and 
$[\Sigma_z,\Sigma_x]= 2i\Sigma_y$. These relations
can be summarized or put into the usual compact notation
namely $\{\Sigma_{i},\Sigma_{j}\}=\delta_{ij}$,
$[\Sigma_{i},\Sigma_{j}]=2i\epsilon_{ijk}\Sigma_{k}$,
$i,\;j=x,\;y,\;z$, we thus have a consistent Pauli
algebra.

Using Eq.~\ref{F8} we may rewrite 
Eq.~\ref{F7} completely in terms of $\Sigma$'s, viz
\begin {eqnarray}
{\cal H}&=&\hbar \nu a^{\dagger}a
+\hbar \frac{\Omega}{2}\Sigma_{z}\nonumber\\
&&-\eta\hbar\frac{\Omega_0}{2}\frac{\Delta}{\Omega}\sin\chi
(\Sigma_{+}a+\Sigma_{-}a^{\dagger}).
\label{F9}
\end{eqnarray}
Comparing the Hamiltonian of Eq.~\ref{F9} with the 
JCH \cite{cir93},
\begin {eqnarray}
{\cal H}=\hbar \omega_{f} a^{\dagger}a
+\frac{1}{2}\omega_0 \sigma_{z}
+\hbar g(\sigma_{+}a+\sigma_{-}a^{\dagger}),
\label{F10}
\end{eqnarray}
we see a direct correspondence. In particular by
making the identifications $\nu \leftrightarrow \omega_f$,
$\Omega \leftrightarrow \omega_0$, and 
$g \leftrightarrow \eta\frac{\Omega_0}{2}\frac{\Delta}{\Omega}
\sin{\chi}$ we immediately see that there is a direct
correspondence between the two-level trapped ion
interacting with a standing-wave light field and the familiar
JCH. The effective coupling 
$\eta\frac{\Omega_0}{2}\frac{\Delta}{\Omega}\sin{\chi}$
is dependent on the Rabi frequency and can be adjusted
in experiments. It is clear that Eq.~\ref{F9} reduces
in the special case of the ion at the node [i.e. setting
$\chi=\pi/2$] to the previous result \cite{cir93}.

	Going back to Eq.~\ref{F4} we see that the second-order 
term in $\eta$ corresponds in the RWA approximation
to the two-photon term in the m-photon JCH 
\cite{buck81,suk84,ala96-1}. The higher-order terms in
$\eta$ will similarly correspond to the multi-photon
terms in the m-photon JCH.

	One can see that the fluorescence spectrum
without including the decay rate will consist of
three peaks as noted in \cite{cir93}, in our
case the three peaks will be centered at
$\nu=0,\;\pm \Omega$ with $\Omega$ given
in \ref{F6}, we note that our $\Omega_0$
corresponds to $\Omega$ in \cite{cir93}.
If the decay rate is included $\Omega$ will be
changed to something like
$\Omega=\sqrt{\Delta^2+\Omega_0^2\cos^2\chi+c~\Gamma^2}$,
where $c$ is a constant. It is claimed that
$c=5$ in \cite{wu97}, however we have not
checked this.

	In conclusion:
\begin{itemize}
\item{}We have treated the interaction of a trapped
ion with standing-wave light field relaxing the
previous condition that the trapping ion remain
on the node of the standing-wave \cite{cir93}.
It has been shown here that the two-level trapped
ion dynamics both off and on the node is 
described by JCH .
\item{}Our analysis relies on a transformation
to the dressed basis, the transformation in turn
is determined by demanding that the Pauli algebra
should hold between the transformed matrices.
\item{}We have also shown/indicated that by
going to higher orders i.e. beyond first order
in Lamb-Dicke parameter $\eta$ we are lead
to the m-photon JCH. 
\end{itemize}

	One could use the well-established field
of ion trapping as a testing ground for strongly
coupled QED because for a trapped ion the coupling
parameters can be varied by laser field strength.
The ability to control/vary the coupling is an
attractive feature of the trapped ion system.\\		
Note added: We recently noticed the work by Wu and Yang \cite{wu97}
which also deals with the extension of \cite{cir93}.
However during the course of their analysis they go
to the interaction picture and in the end drop
the time-dependences. In contrast our approach is
based on maintaining the algebra of Pauli matrices
after the transformation to the ``dressed basis''. And
hence the use of interaction picture in our approach
is not needed in the context of establishing a correspondence
between trapped two-level ion interacting with a standing-wave
light field and JCH. 
\section*{Acknowledgments}
The work of S.~Alam was supported in part by a COE Fellowship 
from the Japanese Ministry of Education and culture and in part
by the JSPS.

\end{document}